\documentclass[12pt]{article}

\usepackage{amsmath,amssymb,amsfonts,amsbsy}
\usepackage{cite}
\usepackage{graphicx}
\usepackage{wrapfig}
\usepackage{epsfig}

\usepackage{bbm} 
\usepackage{bm} 
\usepackage{color}                                                       %
\usepackage{dsfont} 
\usepackage{latexsym} 
\usepackage{lscape} 
\usepackage{mathrsfs} 
\usepackage{morefloats} 
\usepackage{floatflt} 
\usepackage{slashed} 
\usepackage{psfrag}

\usepackage{srcltx}

\DeclareFontFamily{OT1}{mygreek}{}%
\DeclareFontShape{OT1}{mygreek}{m}{n}{<->omsegr}{}%
\DeclareFontShape{OT1}{mygreek}{b}{n}{<->omsegrb}{}%
\DeclareFontShape{OT1}{mygreek}{m}{it}{<->omsegri}{}%
\DeclareFontShape{OT1}{mygreek}{bx}{n}{<->sub * mygreek/b/n}{}%
\DeclareFontShape{OT1}{mygreek}{m}{sl}{<->sub * mygreek/m/it}{}%
\DeclareSymbolFont{Greekrm}{OT1}{mygreek}{m}{n}
\DeclareSymbolFont{Greekbf}{OT1}{mygreek}{b}{n}
\DeclareSymbolFont{Greekit}{OT1}{mygreek}{m}{it}
\DeclareMathSymbol{\omegab}{\mathalpha}{Greekbf}{119}

\begin{document}
\hfill {\small \verb"RUB-TPII-08/2011"}
\addcontentsline{toc}{subsection}{{Spin microscopy with Wilson lines}\\
{\it I.O. Cherednikov}}

\setcounter{section}{0}
\setcounter{subsection}{0}
\setcounter{equation}{0}
\setcounter{figure}{0}
\setcounter{footnote}{0}
\setcounter{table}{0}

\begin{center}
\textbf{SPIN MICROSCOPY WITH ENHANCED WILSON LINES \\ IN THE TMD PARTON DENSITIES\footnote{Presented at the XIV Workshop on High Energy Spin Physics, 20-24 Sept 2011, Dubna, Russia}}

\vspace{5mm}

{I.O.~Cherednikov}$^{\,1,\,2,\,\dag}$, {A.I.~Karanikas}$^{\,3}$ and
N.G.~Stefanis$^{\,4}$

\vspace{5mm}

\begin{small}
  (1) \emph{Departement Fysica, Universiteit Antwerpen, B-2020 Antwerpen, Belgium} \\
  (2) \emph{BLTP, JINR, RU-141980 Dubna, Russia} \\
  (3) \emph{Department of Physics, Nuclear and Particle Physics Section, \\
            Panepistimiopolis, GR-15771 Athens, Greece} \\
  (4) \emph{Institut f\"{u}r Theoretische Physik II,
            Ruhr-Universit\"{a}t Bochum, \\
            D-44780 Bochum, Germany} \\
  $\dag$ \emph{E-mail: igor.cherednikov@ua.ac.be}
\end{small}
\end{center}

\vspace{0.0mm} 

\begin{abstract}
We discuss the possibility of non-minimal gauge invariance of
trans\-ver\-se\--mo\-men\-tum\--de\-pen\-dent parton densities
(TMDs) that allows direct access to the spin degrees of freedom
of fermion fields entering the operator definition of (quark) TMDs.
This is achieved via enhanced Wilson lines that are supplied with
the spin-dependent Pauli term
$\sim F^{\mu\nu}[\gamma_\mu, \gamma_\nu]$,
thus providing an appropriate tool for the ``microscopic''
investigation of the spin and color structure of TMDs.
We show that this generalization leaves the leading-twist
TMD properties unchanged but modifies those of twist three by
contributing to their anomalous dimensions.
We also comment on Collins' recent criticism of our approach.
\end{abstract}

\vspace{7.2mm}

Precise knowledge of the geometrical structure, as well as of the spin
and color properties, of the Wilson lines (gauge links) in the operator
formulation of TMDs is an essential ingredient of the QCD factorization
approach to semi-inclusive hadronic processes \cite{TMD_basic, TMD_INT}.
The path-$[\mathcal{C}]$-dependent non-Abelian gauge links
$
  [y;x|\mathcal{C}]
\! \equiv \!
  \mathcal{P} \exp \!
                  \left[
                  - ig \int_{x[\mathcal{C}]}^{y}
                           dz^\mu A_{\mu}^{a}(z)t^a
                  \right],
$
which ensure the gauge invariance of nonlocal operator products
and correlators, are intimately related to important issues of TMDs,
like the ultraviolet (UV) and rapidity evolution equations, the
generation of $T-$odd effects, the proof or violation of factorization,
etc. \cite{TMD_INT, TMD_gauge_links}.
Different operator definitions of the TMDs can comprise bunches
of longitudinal and transverse gauge links possessing a quite
involved space-time structure, with non-trivial properties in
color space as well
(see, e.g.,  \cite{TMD_gauge_links, New_TMD_Col, New_TMD_PHENO,%
 TMD_LC_trans, CS_all, AP_POL} and further discussions and
references cited therein).
Moreover, the location of the gauge integration contours in the
$(z^+, z^-, \bm z_\perp)$-plane
(in contrast to collinear PDFs, where they belong to a single
lightlike ray and are, therefore, one-dimensional) necessitates
the inclusion of (possible) contributions of
{\it non-minimal} spin-dependent terms, expressed in terms of
enhanced Wilson lines (more below).
The path-dependence, being in some sense ``hidden'' in the case
of collinear PDFs \cite{CS_all}, becomes a key issue in TMDs.
In particular, explicit spin-dependent terms in the gauge links can
create significant effects in lattice simulations
\cite{TMD_INT, TMD_Lattice}, depending on the geometry of the
integration paths, and may also affect the TMD-factorization
properties \cite{TMD_gauge_links}.

To this end, we introduced in \cite{CKS10} an enhanced gauge
link, denoted by $[[...]]$, which contains the Pauli term proportional
to the gluon strength tensor
$
\sim
 F_{\mu\nu}^{a} J_{\mu\nu}
=
 (1/4)F_{\mu\nu}^{a} [\gamma_\mu, \gamma_\nu].
$
This is the simplest example to realize a direct product of two
orthogonal ``spaces'': The first ``space'' is the color one, with the
{\it minimal} Wilson lines in the fundamental or adjoint representation
of $SU(3)_{\rm c}$.
In the second ``space'', the spin correlations are generated by the
Pauli terms \cite{CKS10}.
The spin-dependent terms yield next-to-leading-order twist
effects with respect to the spin-``blind'' ones, as it follows from
usual power-counting.

We discuss below, the main results of our study of the
renormalization-group properties of the TMD distribution functions with
enhanced gauge-link insertions \cite{CKS10}, focusing on
the UV properties of the ``quark-in-a-quark'' TMD.
According to our generalized concept of gauge invariance, the
{\it unsubtracted} distribution function of a quark with momentum $k$
and flavor $a$ in a quark with momentum $p$ reads
\begin{eqnarray}
& &
{\cal F}_{a}^{\Gamma}(x, \bm{k}_{\perp})
 = {}
   \frac{1}{2} {\rm Tr} \! \int\! dk^-
   \int \! \frac{d^4 \xi }{(2\pi)^4}\,
   {\rm e}^{- i k \cdot \xi}
\langle  p, s \ |\bar \psi_a (\xi) [[\xi^-, \bm{\xi}_{\perp};
   \infty^-, \bm{\xi}_{\perp}]]^\dagger \nonumber \\
& &
 \times
 [[\infty^-, \bm{\xi}_{\perp};
   \infty^-, \bm{\infty}_{\perp}]]^\dagger \Gamma
\left.
   [[\infty^-, \bm{\infty}_{\perp};
   \infty^-, \mathbf{0}_{\perp}]]
   [[\infty^-, \mathbf{0}_{\perp};
   0^-, \mathbf{0}_{\perp}]]
\psi_a (0) | p, s \right \rangle \ ,
\vspace{-1cm}
\label{eq:TMD-PDF}
\end{eqnarray}
where $\Gamma$ stands for the Dirac structure constructed from
one or several $\gamma$-matrices.
The matrix elements interpolate between the one-fermion states
with momentum $p$ and spin $s$:
$| p, s \rangle$.
In the tree-approximation one has
\begin{equation}
  {\cal F}^{{\Gamma} (0)}(x, \mbox{\boldmath$k_\perp$})
=
  \frac{1}{2}
  {\rm Tr} \left[ (\hat p + m)
  \left(1 + \gamma_5 {\hat s} \right) \ \Gamma
           \right] \
  \delta(p^+ - xp^+) \delta^{(2)} (\bm k_\perp)
  \ .
\end{equation}
For the unpolarized TMD PDF with
$\Gamma = \gamma^+$,
one obtains the (twist-two) result
\begin{equation}
  {\cal F}^{\gamma^+ (0)}(x, \mbox{\boldmath$k_\perp$})
\! = \!
  \frac{1}{2}
  {\rm Tr} \left[ (\hat p + m)
  \left(1 + \gamma_5 {\hat s} \right) \gamma^+
           \right]
  \delta(p^+ - xp^+) \delta^{(2)} (\bm k_\perp)
\! = \!
  \delta(1-x) \delta^{(2)} (\bm k_\perp)
  \ .
\end{equation}
The helicity and the transversity distributions read, respectively,
\begin{equation}
  {\cal F}^{\gamma^+\gamma_5 (0)}(x, \mbox{\boldmath$k_\perp$})
=
  \delta(1-x) \delta^{(2)} (\bm k_\perp) \cdot \lambda
  \ ,
\
  {\cal F}^{i \sigma^{i+}\gamma_5 (0)}(x, \mbox{\boldmath$k_\perp$})
=
  \delta(1-x)
  \delta^{(2)} (\bm k_\perp)\! \cdot\! \mbox{\boldmath$s_{\perp}^{i}$} \ ,
\end{equation}
where $\lambda$ is the helicity and $\mbox{\boldmath$s_{\perp}^{i}$}$
is the transverse spin of parton $i$.
Note that the above normalization conditions can only
be obtained within the {quantization procedure in the light-cone gauge},
where the (minimal) longitudinal Wilson lines vanish and the equal-time
canonical commutation relations for the quark creation and annihilation
operators
$\{a^\dag (k, \lambda), a(k, \lambda)\}$
are consistent with the {\it parton-number interpretation} of the TMD
in the tree-approximation (see \cite{ChSt_Wilson} for more):
$
 {\cal F}^{(0)} (x, {\bm k}_\perp)
\sim
 \langle  p  | a^\dag(k^+, \bm k_\perp;
 \lambda)\ a(k^+, \bm k_\perp; \lambda) | p \rangle.
$
In line with the above explanations, we define a generic enhanced gauge
link evaluated along some fixed but else arbitrary direction $w$ from
zero to infinity according to
\begin{equation}
  [[\infty; 0]]
 =
  \mathcal{P}
  \exp
      \left[
            - ig \int_{0}^{\infty} d\sigma \ w_{\mu} \
                 A_{a}^{\mu}(w \sigma)t^a
           - i g \int_{0}^{\infty} d\sigma  \
                 J_{\mu\nu} F_{a}^{\mu\nu}(w \sigma)t^a
      \right] \ ,
\label{eq:lightlike-link}
\end{equation}
where the four-vector $w$ may be longitudinal (light-like)
$w_L = n^-$, or transverse
$w_T = (0^+, 0^-, \bm l_\perp)$.
The enhanced Wilson lines (\ref{eq:lightlike-link}) significantly
enlarge the gauge-invariant formalism of quark and gluon operators
entering the TMD correlators.

To investigate the structure of the UV singularities in the leading
$\alpha_s$-order, we evaluate all graphs contributing to this
order given in \cite{CKS10}, where one can also find the technical
details and the appropriate Feynman rules.
Note that there are two different perturbative expansions in
the generalized TMD given by (\ref{eq:TMD-PDF}): one stems from the
Heisenberg quark field operators, i.e.,
$
  \psi_a(\xi)
 =
  {\rm e}^{- ig\left[ \int\! d\eta \ \bar \psi \hat {\cal A} \psi
               \right]} \
  \psi_a^{\rm free} (\xi),
$  \
$
  \int\! dx \ \bar \psi \hat {\cal A} \psi \ \equiv
  \int\! d^4 x \ \bar \psi (x) \gamma_\mu
  \psi (x) {\cal A}^\mu (x).
$
The other originates from the evaluation of the product of the
enhanced gauge links up to $\mathcal{O}(g^2)$.
Applying the light-cone gauge
$A^+  = 0$) one has
\begin{equation}
  [[\infty^-, \mbox{\boldmath$\infty_\perp$};
    \infty^-, \mbox{\boldmath$0_\perp$}
  ]]
\cdot
  [[\infty^-, \mbox{\boldmath$0_\perp$};
    0^-, \mbox{\boldmath$0_\perp$}
  ]]
=
   1 - i g \left(
               \mathcal{U}_{1} + \mathcal{U}_{2} + \mathcal{U}_{3}
         \right)
    - g^2 \left(
                \mathcal{U}_{4} + \mathcal{U}_{5}
                + \ldots \mathcal{U}_{10}
          \right) \ ,
\label{eq:gauge-links-product}
\end{equation}
where the individual contributions $\mathcal{U}_i$ have to
be contracted with themselves as well as with corresponding terms
in the Heisenberg field operators.

The singularity structure of the twist-two TMD with the
Dirac structures
$
 \Gamma_{\rm tw\,-\,2}
=
 \{ \gamma^+, \gamma^+\gamma^5, i\sigma^{i+}\gamma^5 \}
$,
cancel by the Hermitean conjugated (mirror) diagrams, in contrast to
the twist-three TMDs
(e.g., $\Gamma_{\rm tw\,-\,3} = \gamma^i$) which receive non-trivial
UV divergent contributions from the Pauli term, like
\begin{equation}
  \Gamma_{\rm tw\,-\,3} \langle \mathbf{A}^{\perp} F^- \rangle
  + \langle \mathbf{A}^{\perp} F^- \rangle \Gamma_{\rm tw\,-\,3}
=
  - C_{\rm F} \ \frac{1}{4\pi} [\gamma^+, \gamma^-]
  \ \Gamma (\varepsilon) \
  \left( 4\pi \frac{\mu^2}{\lambda^2}\right)^\varepsilon \ .
\label{eq:uv_singular}
\end{equation}
Here,
$\langle \mathbf{A}^{\perp} F^- \rangle$
denotes the result stemming from the cross-talk between the minimal
transverse gauge link and the enhanced longitudinal gauge link
containing a Pauli term.
In order to render the TMD singularity-free, one has to handle
the overlapping UV and rapidity divergences induced by the gluon
propagator in the lightcone gauge.
To this end, we refurbished in \cite{CS_all} the untruncated definition
in Eq.\ \ref{eq:TMD-PDF} by a soft renormalization factor along a
particular gauge contour going off the lightcone.
This soft factor takes care of the overlapping UV and
infrared (rapidity) divergences which cannot be
regularized dimensionally, as in the case of purely longitudinal
gauge links---see \cite{Ste83} and references cited therein.

Recently, Collins \cite{New_TMD_Col} questioned the validity
of this definition and proposed another one.
He argues that the gluon propagator in the lightcone gauge subject to
the Mandelstam-Leibbrandt (ML) boundary prescription,
$D_{\rm ML}^{\mu\nu}$,
is not transverse, i.e., $n_\mu D_{\rm ML}^{\mu\nu}\neq 0$.
The propagator displayed by Collins as Eq. (15) in \cite{New_TMD_Col}
is \textit{not} the ML one but the result of using the Principal-Value
prescription.
This propagator, as well as the Retarded and the Advanced one, are
indeed not transverse.
In contrast, the \textit{correct} ML propagator (see last
entry in \cite{CS_all}) \textit{is} transverse and the soft factor
reduces to unity.
The second argument by Collins is that the graphs shown in Eq.\ (16)
in \cite{New_TMD_Col} give uncanceled rapidity divergences.
If the displayed graphs are to be evaluated in the lightcone gauge,
as used in our works in \cite{CS_all} and in \cite{CKS10}, then they
both vanish.
In a general covariant gauge, these graphs contribute singularities
that are indispensable in order to cancel those singular terms,
induced by the gluon propagator, which contain the gauge parameter.
There are no surviving singularities.

In conclusion, we discussed a new operator formulation of
gauge-invariant TMDs which provides direct access to the spin degrees
of freedom of the partonic fields by means of the Pauli term in the
gauge links, hence allowing a microscopic analysis of the spin-color
structure of TMDs relevant for phenomenology.

\paragraph{Acknowledgements}
The first author (I.O.C.) thanks the Organizers for the warm
hospitality and exciting atmosphere during the conference.

\end{document}